\documentclass[12pt]{article}
\usepackage{a4}
\usepackage{amsfonts}
\usepackage{amsmath,amssymb}

\numberwithin{equation}{section} 

\makeatletter
\def\un#1{\relax\ifmmode\@@underline#1\else
        $\@@underline{\hbox{#1}}$\relax\fi}
\makeatother

\let\du=\du                     

\def\a{\alpha}
\def\b{\beta}
\def\c{\chi}
\def\d{\delta}
\def\e{\epsilon}
\def\f{\phi}
\def\g{\gamma}
\def\h{\eta}

\def\j{\psi}

\def\l{\lambda}
\def\m{\mu}
\def\n{\nu}
\def\o{\omega}

\def\r{\rho}
\def\s{\sigma}

\def\x{\xi}
\def\z{\zeta}

\def\L{\Lambda}
\def\O{\Omega}

\def\ve{\varepsilon}

\def\cf{{\cal F}}

\def\cl{{\cal L}}

\def\cv{{\cal V}}

\def\bo{{\raise-.3ex\hbox{\large$\Box$}}}               
\def\pa{\partial}                                       
\def\TH{{\raise.2ex\hbox{$\displaystyle \bigodot$}\mskip-4.7mu \llap H \;}}
\def\face{{\raise.2ex\hbox{$\displaystyle \bigodot$}\mskip-2.2mu \llap {$\ddot
        \smile$}}}                                      

\def\Tilde#1{\widetilde{#1}}                    
\def\Bar#1{\overline{#1}}                       
\def\leftrightarrowfill{$\mathsurround=0pt \mathord\leftarrow \mkern-6mu
        \cleaders\hbox{$\mkern-2mu \mathord- \mkern-2mu$}\hfill
        \mkern-6mu \mathord\rightarrow$}
\def\dvec#1{\vbox{\ialign{##\crcr
        \leftrightarrowfill\crcr\noalign{\kern-1pt\nointerlineskip}
        $\hfil\displaystyle{#1}\hfil$\crcr}}}           

\def\frac#1#2{{\textstyle{#1\over\vphantom2\smash{\raise.20ex
        \hbox{$\scriptstyle{#2}$}}}}}                   
\def\half{\frac12}                                        
\def\sfrac#1#2{{\vphantom1\smash{\lower.5ex\hbox{\small$#1$}}\over
        \vphantom1\smash{\raise.4ex\hbox{\small$#2$}}}} 
\def\bfrac#1#2{{\vphantom1\smash{\lower.5ex\hbox{$#1$}}\over
        \vphantom1\smash{\raise.3ex\hbox{$#2$}}}}       
\def\afrac#1#2{{\vphantom1\smash{\lower.5ex\hbox{$#1$}}\over#2}}    

\def\[{\lfloor{\hskip 0.35pt}\!\!\!\lceil}
\def\]{\rfloor{\hskip 0.35pt}\!\!\!\rceil}

\def\du#1#2{_{#1}{}^{#2}}

\def\un{\underline}
\def\fracmm#1#2{{{#1}\over{#2}}}

\def\low#1{{\raise -3pt\hbox{${\hskip 0.75pt}\!_{#1}$}}}

\def\Tilde#1{{\widetilde{#1}}\hskip 0.015in}

\newskip\humongous \humongous=0pt plus 1000pt minus 1000pt

\newif\ifdtup

\newcommand{\be}{\begin{equation}}
\newcommand{\ee}{\end{equation}}
\newcommand{\nbe}{\begin{equation*}}
\newcommand{\nee}{\end{equation*}}

\begin{document}

\thispagestyle{empty}

{\hbox to\hsize{
\vbox{\noindent July 2007 \hfill }}}

\noindent
\vskip1.3cm
\begin{center}

{\Large\bf $N=1/2$~ Supergravity with Matter in Four Euclidean 
Dimensions~\footnote{Supported in part by the Japanese Society for Promotion 
of Science (JSPS)}}
\vglue.2in

Tomoya Hatanaka~\footnote{Email address: hatanaka-tomoya@c.metro-u.ac.jp}
and Sergei V. Ketov~\footnote{Email address: ketov@phys.metro-u.ac.jp}
\vglue.1in

{\it Department of Physics\\
     Tokyo Metropolitan University\\
     1--1 Minami-osawa, Hachioji-shi\\
     Tokyo 192--0397, Japan}
\end{center}
\vglue.2in
\begin{center}
{\Large\bf Abstract}
\end{center}

\noindent

An $N=1/2$ supergravity in four Euclidean spacetime dimensions, coupled to both
vector- and scalar-multiplet matter, is constructed for the first time. We 
begin with the standard $N=(1,1)$ conformally extended supergravity in four 
Euclidean dimensions, and freeze out the graviphoton field strength to an 
arbitrary (fixed) self-dual field (the so-called $C$-deformation). Though a 
consistency of such procedure with local supersymmetry is not guaranteed, we 
find a simple consistent set of algebraic constraints that reduce the local 
supersymmetry by 3/4 and eliminate the corresponding gravitini. The final field
theory (after the superconformal gauge-fixing) has the residual local 
$N=(0,\frac{1}{2})$ or just $N=1/2$ supersymmetry with only one chiral 
gravitino as the corresponding gauge field. Our theory is not 
`Lorentz'-invariant because of the non-vanishing self-dual graviphoton vacuum 
expectation value, which is common to the $C$-deformed $N=1/2$ rigidly 
supersymmetric field theories constructed in a non-anticommutative superspace.
 
\newpage

\section{Introduction}
\vglue.2in
A construction of new supergravity theories is apparently complete after 
a lot of work done in the past --- see e.g., refs.~\cite{rev1, rev2}. However,
it is merely apparent, because some recent developments in field theory, 
strings and gravity  offer new opportunities for even further generalizations 
of supergravities, by relaxing some of the symmetry requirements. It offers 
new perspectives to various physical applications, such as (i) partial 
supersymmetry breaking, and (ii) brane supersymmetry reduction.

One of the recent developments is a noncommutative gravity. Though the
idea of replacing the ordinary field product by the noncommutative Moyal 
(star) product is not new \cite{ncgr}, its implementation is not unique, while 
it often leads to a complexification of the metric and the appearance of 
ghosts (see, however, refs.~\cite{cure} for possible cures). The appearance
of infinitely many interaction vertices with unlimited powers of momenta is 
the necessary feature of those noncommutative gravity models.

String theory can teach us more about noncommutative gravity (see e.g., 
ref.~\cite{cern}), as well as about supersymmetry \cite{ov}. In particular, 
as was observed by Ooguri and Vafa \cite{ov}, the superworldvolume of a 
supersymmetric D-brane in a constant Ramond-Ramond type flux gives rise to
the remarkable new structure in the corresponding superspace, which is now
called Non-AntiCommutativity (NAC). It means that the fermionic superspace 
coordinates are no longer Grassmann (i.e. they no longer anti-commute), but 
satisfy a Clifford algebra. In other words, the impact of the Ramond-Ramond 
flux on the D-brane dynamics can be described by the non-anticommutativity in 
the D-brane superworldvolume. 

As regards a D3-brane, a 10-dimensional (self-dual) five-form flux upon 
compactification to four dimensions gives rise to the (self-dual) graviphoton 
flux in the D3-brane 4-dimensional worldvolume \cite{ov}. In its turn, the 
non-anticommutativity in superspace can be described by the (Moyal-Weyl type) 
non-anticommutative star product amongst superfields. It results in a 
construction of the NAC deformed supersymmetric field theories with partially 
broken supersymmetry, pioneered by Seiberg \cite{sei}, in four Euclidean 
dimensions. Unlike bosonic noncommutativity, the NAC supersymetric field 
theories usually have only a limited number of new interaction terms, without 
higher derivatives, while their 
Lagrangians can often be written down in closed form.  As a matter of
fact, all recent studies of the NAC supersymmetric field theories, following
ref.~\cite{sei}, were limited to {\it rigid\/} supersymmetry, i.e. without 
gravity or supergravity (see e.g., ref.~\cite{our} and references therein). 

Given the relation between a non-anticommutativity and a non-vanishing 
(self-dual) vacuum expectation value of a graviphoton in four dimensions, 
it is quite natural to explore further possiblities for a construction of new
supergravities, by freezing a graviphoton field in an extended supergravity
theory (with matter). The minimally extended Poincar\'e supergravity in four 
dimensions, that has a graviphoton as the superpartner of a graviton, is the 
$N=(1,1)$ or just $N=2$ supergravity. The structure of $N=2$ supergravity with
matter was given in detail in refs.~\cite{hol,italy}.

In our earlier paper \cite{hat}, a toy model of the four-dimensional $N=1/2$ 
supergravity with a fixed self-dual graviphoton expectation value  was 
constructed by freezing out the graviphoton field strength in the 
standard $N=(1,1)$ extended supergravity with two non-chiral gravitini 
\cite{fn}.  Our supergravity model \cite{hat} has local $N=(0,\frac{1}{2})$ 
supersymmetry. Consistency of the model \cite{hat} requires the expectation
value of the graviphoton field strength to be equal  to the self-dual 
(bilinear) gravitino condensate. 

An extension of the construction \cite{hat} to a matter-coupled $N=1/2$ 
supergravity is not automatic since more consistency conditions have to
be satisfied. In this paper we report our results of such construction, by
presenting the Lagrangian and the local supersymmetry transformation laws
of the $N=1/2$ supergravity in four Euclidean dimensions, coupled to vector
supermultiplets and scalar supermultiplets, with all the fermionic terms
included.

Our paper is organized as follows: in Sec.~2 we briefly describe the contents 
of $N=2$ conformal supergravity multiplets in the $N=2$ superconformal tensor 
calculus \cite{hol}. In Sec.~3 we introduce the 2-component notation for 
spinors in four Euclidean dimensions. In Sec.~4 we describe our way of freezing
 of a graviphoton field, by imposing a self-duality condition on the 
graviphoton field strength, and then studying its consistency and the surviving
 symmetries.~\footnote{As regards a consistent reduction of the $N=2$ 
matter-coupled  supergravity to $N=1$ matter-coupled supergravity, including 
all fermionic terms, see ref.~\cite{2to1}.}  The residual superconformal 
transformations are given in Sec.~5. The superconformal gauge-fixing and 
elimination of the auxiliary fields are discussed in Sec.~6. Our Lagrangian of 
the $N=1/2$ Euclidean supergravity with matter is given in Sec.~7. Sec.~8 is 
our Conclusion. Three Appendices are devoted to further notation and some 
technical details. The notation \cite{hol} for hypermultiplets is briefly 
summarized in Appendix A. In Appendix B we collect the $N=2$ superconformal 
transformation laws \cite{hol} which is the starting point of our construction.
  In Appendix C we quote the so-called decomposition rules \cite{hol} needed in
 passing from the conformal supergravity to the `Poincar\'e' supergravity. 

\section{$N=2$ supergravity field components}

The last paper of ref.~\cite{hol} is the pre-requisite to our construction. So,
 instead of copying the equations of ref.~\cite{hol} here, we merely review the
 methods of ref.~\cite{hol}, and concentrate on the differences between our 
construction and that of ref.~\cite{hol}.

First, our construction can only be defined in {\it Euclidean} four-dimensions,
 not in Minkowski spacetime as in ref.~\cite{hol}. As is well known, the c
hange of signature has important implications on the structure of field 
representations, especially on spinors. For instance, minimal spinor 
representations in Minkowski signature are given by real (Majorana) spinors or 
complex chiral spinors, while the chiral and anti-chiral parts of a Majorana 
spinor are related by complex conjugation. Accordingly, a number of 
supersymmetries in Minkowski spacetime is measured by a number of Majorana 
supercharges. In four Euclidean dimensions, Majorana spinors do not exist 
\cite{tnw}, whereas the chiral and anti-chiral spinors are independent. Hence, 
the numbers of left and right (chiral and anti-chiral) Euclidean supercharges 
need not be the same,  while the minimal choice is obviously given by one 
chiral or anti-chiral supercharge. We call it $(1/2,0)$ or $(0,1/2)$ susy,
 respectively, or simply $N=1/2$ supersymmetry.
 
Second, in order to make the chiral supersymmetry manifest, we are going to use
 the 2-component notation for spinors, which is best suitable for our purposes.
 So we rewrite the results of ref.~\cite{hol} obtained in the 4-component 
spinor notation, to the 2-component notation in four Euclidean dimensions 
(see Sec.~3). 

The $N=2$ superconformal tensor calculus gives us a systematic method for 
constructing the $N=2$ super-conformal and super-Poincar\'e-invariant
 Lagrangians and the $N=2$ transformation laws (see e.g., ref.~\cite{apro} for 
a review.) It provides us with (i) the off-shell $N=2$ supermultiplets, as the 
representations of $N=2$ local superconformal algebra, together with the 
transformation laws of their field components, which form a closed algebra, 
(ii) the multiplication rules for a construction of new representations, and 
(iii) the density formulas describing the superconformal invariants.

In order to get the super-Poincar\'e Lagrangian and the transformation laws, 
one has to fix the truly superconformal symmetries, while keeping the 
super-Poincar\'e ones. It is often called `gauge fixing'. The gauge fixing 
conditions give rise to the decomposition laws relating the truly 
superconformal transformation parameters to the super-Poincar\'e transformation
 parameters --- see refs.~\cite{hol,apro} for details.

The off-shell $N=2$ superconformal multiplets are given by \cite{hol}
\begin{itemize}
\item a {\it Weyl} multiplet,
\item a {\it vector} multiplet, 
\item a {\it hyper}multiplet. 
\end{itemize}

The $N=2$ {\it Weyl} multiplet has $24+24$ independent field components:
\be
(e_\m{}^a, \j_\m^i, b_\m, A_\m, \cv_\m{}^i{}_j, T_{ab}{}^{ij}, \c^i, D)
\ee
where $e_\m^a$ is vierbein, and the gravitino doublet $\j_\m^i$ is the gauge 
field of $N=2$ local supersymmetry. The gauge fields of other superconformal 
symmetries are $b_\m$ for dilatations, $A_\m$ for chiral $U(1)$ rotations, and 
$\cv_\m{}^i{}_j$ for chiral $SU(2)$ rotations.  We also need the auxiliary 
fields: the bosonic tensor $T_{ab}^{ij}$ and the real scalar $D$, and a  
fermionic (spinor) doublet $\c^i$. 

Only $e_\m^a$ and $\j_\m^i$ are going to represent physical degrees of freedom,
 the $\cv_\m{}^i{}_j$ is the antihermitian traceless matrix in its $SU(2)$ 
indices $i,j$, while $T_{ab}^{ij}$ is the real tensor 
antisymmetric in its both $SU(2)$ and `Lorentz' index pairs. 

An $N=2$ {\it vector} multiplet has $8+8$ independent field components:
\be
(X, \O_i, W_\m, Y_{ij})
\ee
where $X$ is a complex scalar, $\O_i$ is a spinor doublet, $W_\m$ is a vector 
gauge field, and $Y_{ij}$ is 
a real $SU(2)$ auxiliary triplet. 

We consider the vector gauge fields to be Lie-algebra valued, with the 
hermitean generators $t_I$ obeying an algebra
\be
	[t_I, t_J]=f_{IJ}{}^K t_K
\ee
where we have introduced the Lie algebra structure constants $f_{IJ}{}^K$. So, 
the vector multiplet components
 carry the extra (gauge) index, $I,J,\dots =0,1,\dots,n$. We choose $I=0$ for 
a graviphoton (abelian)
gauge field, so we also define $\Tilde{I}, \Tilde{J}, \dots =1,2, \dots ,n$. 

The {\it hyper}multiplet physical fields are given by
\be
(A_i{}^\a,\z^\a)
\ee
where $A_i$ is a scalar doublet and $\z$ is a complex spinor. The 
hypermultiplets are supposed to belong to a 
representation of the non-abelian gauge group. The index \(i=1,2\) is 
associated to the $SU(2)$ automorphism 
group of the $N=2$ supersymmetry algebra, whereas the index 
\(\a=1,2,\cdots 2r\) is the representation index 
with respect to the non-abelian gauge group. See Appendix A for more about the 
hypermultiplet notation.

The $N=2$ superconformal transformation laws in the 2-component Euclidean
notation are collected in Appendix B.

A consistent Wick rotation of a field theory with fermions from four Minkowski 
dimensions to four Euclidean 
dimensions is described in detail in ref.~\cite{tnw}. In ref.~\cite{hol} the 
spacetime signature
$(+++-)$ was used, which is now going to be Wick-rotated to $(++++)$ by setting
 $x_4=it$. As regards the
vector gauge fields, it implies  $A_\m \to (\vec{A}{}^E, iA_4^E)$. As is argued
 in ref.~\cite{tnw}, 
one should change $\g_4 \to i\g_{E}^5$, and use gamma matrices and a charge 
conjugation matrix in four
Euclidean dimensions (see e.g., an Appendix in ref.~\cite{rev1}).~\footnote{In 
ref.~\cite{tnw} the Dirac
conjugation includes a factor of $i$, while the Lagrangian excludes a factor of
 $i$, but we are going
to use the opposite notation.} So our definition of the Dirac conjugation is
\be
\Bar{\l}{}^i=(\l_i)^\dagger i\g_E^5.
\ee
For instance, the Majorana condition is modified as follows: 
\be
(\Bar{\l}{}^i)^T C_{E}=(\l_i)^\dagger i\g_E^5
\ee
It is worth mentioning that this condition is {\it not } a reality 
condition for spinors. 
Nevertheless, we can still use this condition for constructing the Euclidean 
version of a given supergravity
theory. To avoid confusion, we sometimes append a script (E) for Euclidean 
fields or matrices, and a script
(M) for their Minkowski counterparts.

\section{Euclidean 2-component spinor notation}

We use lower case Greek letters for curved space vector indices, $\m, \n, 
\dots = 1, 2, 3, 4$,  
lower case Latin indices for flat (tangent) space vector indices, $a, b, 
\dots = 1, 2, 3, 4$, and 
capital Latin letters for (anti)chiral spinor indices (dotted or undotted), 
$A, B, \dots = 1, 2$. 

Gamma matrices in four Euclidean dimensions satisfy an algebra
\begin{equation}
	\{\g_{a}, \g_{b}\}=2\d_{ab}, 
\quad
	\{\g_{5}, \g_{a}\}=0
\end{equation}
An explicit representation of the Euclidean gamma matrices is as follows 
\cite{rev1}:
\begin{equation}
	\g_{k}
	=
	\left(
		\begin{array}{cc}
			0                & -i\s_k^{A\dot{B}} \\
			i\s_{k\dot{A}B} & 0      \\
		\end{array}
	\right),
\quad
	\g_{4}
	=
	\left(
		\begin{array}{cc}
			0 & 1  \\
			1 & 0 \\
		\end{array}
	\right),
\quad
	\g_{5 E}
	=
	\left(
		\begin{array}{cc}
			1 & 0  \\
			0 & -1 \\
		\end{array}
	\right),
\end{equation}
where $\s_k$ are Pauli matrices, $k=1,2,3$.  In addition, we define the 
matrices
\begin{equation}
	\s^{ab}
	=
	\frac{1}{4}
	\left(
		\begin{array}{cc}
			(\s^a\s^b-\s^b\s^a)^A{}_C  & 0                \\
			0 & (\s^\a\s^b-\s^b\s^a)_{\dot{A}}{}^{\dot{C}} \\
		\end{array}
	\right)
	=:
	\left(
		\begin{array}{cc}
			\s^{ab}{}^A{}_C  & 0                \\
			0 & \s^{ab}{}_{\dot{A}}{}^{\dot{C}} \\
		\end{array}
	\right)
\end{equation}
which are anti-hermitean, 
\begin{equation}
	(\s_{ab})^\dagger=-\s_{ba}~~,
\end{equation}
in terms of their self-dual and anti-self-dual combinations,
\begin{equation}
	\frac{1}{2}\ve^{abcd}\s_{cd}{}_{\dot{A}}{}^{\dot{B}}
	=
	\s^{ab}{}_{\dot{A}}{}^{\dot{B}}
\quad
	{\rm and}
\qquad
	\frac{1}{2}\ve^{abcd}\s_{cd}{}^A{}_B
	=
	-\s^{ab}{}^A{}_B
\quad
\label{(anti)selfdual}
\end{equation}

The Euclidean charge conjugation matrix is given by
\begin{equation}
	C
	=
	\g_{4 }\g_{2 }
	=
	\left(
		\begin{array}{cc}
			i\s_2 & 0      \\
			0     & -i\s_2 \\
		\end{array}
	\right)
	=
	\left(
		\begin{array}{cc}
			\e_{AB} & 0             \\
			0 & \e^{\dot{A}\dot{B}} \\
		\end{array}
	\right)
\end{equation}
so that 
\begin{equation}
	C=-C^T~,
\quad
	C\g^a C^{-1}=-(\g^a)^T~,
\end{equation}
\begin{equation}
	C^\dagger=C^{T*}=-C^*~,
\quad
	C^*
	=
	\left(
		\begin{array}{cc}
			\e^{AB} & 0       \\
			0       & \e_{\dot{A}\dot{B}} \\
		\end{array}
	\right)~,
\end{equation}
\begin{equation}
	C^* C=CC^*=-I
\end{equation}

When changing the notation \cite{hol} by representing all the 4-component 
spinors in terms of their 
2-component constituents in Euclidean space, it is important to observe the 
chirality of each 
4-component spinor given by the position (up or down) of its $SU(2)$ index 
$(i,j=1,2)$, e.g., 
\begin{equation}
	\O_{i}^I
	=:
	\left(
		\begin{array}{c}
				\O_i^{I B} \\
				0          \\
		\end{array}
	\right),
\quad
	\O^{iI}
	=:
	\left(
		\begin{array}{c}
				0                           \\
				\Bar{\O}{}^{iI}{}_{\dot{B}} \\
		\end{array}
	\right)~,
\end{equation}
\begin{equation}
	\Bar{\O}{}^{iI}_{\rm M}
\to
	(\O^{iI}_{\rm E}){}^{T}C_{\rm E}
	=:
	\left(
		0, -\Bar{\O}{}^{iI\dot{A}}
	\right)~,
\quad
	\Bar{\O}{}_{{\rm M}i}^I
\to
	(\O^{I}_{{\rm E}i}){}^{T}C_{\rm E}
	=:
	\left(
		\O^I_{Ai}, 0
	\right).
\end{equation}
and
\begin{equation}
	\j^i_{\m }
	=:
	\left(
		\begin{array}{c}
				\j_\m^{iB} \\
				0          \\
		\end{array}
	\right),
\quad
	\j_{\m i }
	=:
	\left(
		\begin{array}{c}
				0                           \\
				\Bar{\j}{}_{\m i \dot{B}} \\
		\end{array}
	\right)~,
\end{equation}
\begin{equation}
	\Bar{\j}{}_{\m {\rm M}}^i
\to
	(\j_{\m {\rm E}}^i){}^{T}C_{\rm E}
	=:
	\left(
		\j^{i}_{\m A}, 0
	\right),
\quad
	\Bar{\j}{}_{\m i{\rm M}}
\to
	(\j_{\m i {\rm E}}){}^{T}C_{\rm E}
	=:
	\left(
		0, -\Bar{\j}{}_{\m i}^{\dot{A}}
	\right)
\end{equation}
In order to avoid double counting, their complex conjugates are given by
\begin{equation}
	(\O^{iI})^*
	=
	\left(
	\begin{array}{c}
		-i\O_{iA}^I	\\
		0
	\end{array}
	\right),
\quad
	(\O^{iI})^\dagger
	=
	(-i\O_{iA}^I, 0)~,
\end{equation}
\begin{equation}
	(\O^{I}_i)^*
	=
	\left(
	\begin{array}{c}
		0	\\
		-i\Bar{\O}^{iI{\dot{A}}}
	\end{array}
	\right),
\quad
	(\O^{I}_i)^\dagger
	=
	(0, -i\Bar{\O}^{iI{\dot{A}}})~,
\end{equation}

\begin{equation}
	(\j_\m^i)^*
	=
	\left(
	\begin{array}{c}
		0	\\
		-i\Bar{\j}{}_{\m i}^{\dot{A}}
	\end{array}
	\right),
\quad
	(\j_{\m}^i)^\dagger
	=
	(0, -i\Bar{\j}{}_{\m i}^{\dot{A}})~,
\end{equation}
\begin{equation}
	(\j_{\m i})^*
	=
	\left(
	\begin{array}{c}
		-i\j_{\m A}^i	\\
		0
	\end{array}
	\right),
\quad
	(\j_{\m i})^\dagger
	=
	(-i\j_{\m A}^i, 0)
\end{equation}

The $SU(2)$ indices are contracted by \(\ve_{ij}\) and \(\ve^{ij}\),
\begin{equation}
	\ve_{ij}\ve^{ij} = 2~,
\quad
	\ve_{12}=-\ve_{21}=\ve^{12}=-\ve^{21}=1~,
\quad
	(\ve_{ij})^\dagger=\ve^{ik}\ve^{jl}\ve_{kl}=\ve^{ij}
\end{equation}

Two-component spinor indices are contracted as follows:
\begin{equation}
	\j_A=\j^B\e_{BA}~,
\quad	
	\j^A=\e^{AB}\j_B~,
\quad
	\Bar{\j}_{\dot{A}}=\Bar{\j}^{\dot{B}}\e_{\dot{B}\dot{A}}~,
\quad
	\Bar{\j}^{\dot{A}}=\e^{\dot{A}\dot{B}}\j_{\dot{B}}~,
\end{equation}
\begin{equation}
	\s_{a \dot{D}C}=\s_a^{A\dot{B}}\e_{AC}\e_{\dot{B}\dot{D}}~,
\quad
	\s_a^{A\dot{B}}=\e^{\dot{B}\dot{D}}\e^{AC}\s_{a\dot{D}C}~,
\end{equation}
where we have
\begin{equation}
	\e_{AB}=-\e_{BA}~,
\quad
	\e^{AB}\e_{BC}=-\d^A{}_C~,
\quad
	\e_{12}=\e^{12}=-\e_{\dot{1}\dot{2}}=-\e^{\dot{1}\dot{2}}=1
\end{equation}
We use the following book-keeping notation:
\begin{equation}
\begin{split}
&	\l\s_a\Bar{\c}
	\equiv
	\l_A\s_a^{A\dot{B}}\Bar{\c}_{\dot{B}},
\quad
	\l\s^a\s^b\c
	\equiv
	\l_A\s^{a A\dot{B}}\s^b_{\dot{B}C}\c^C,
\quad
	\l\c\equiv\l^D\c_D~,
\\
&	\Bar{\l}\s_a\c
	\equiv
	\Bar{\l}^{\dot{A}}\s_{a\dot{A}B}\c^B,
\quad
	\Bar{\l}\s^a\s^b\Bar{\c}
	\equiv
	\Bar{\l}^{\dot{A}}
	\s^a_{\dot{A}B}\s^{b B\dot{C}}
	\Bar{\c}_{\dot{C}},
\quad
	\Bar{\l}\Bar{\c}
	\equiv
	\Bar{\l}^{\dot{D}}\Bar{\c}_{\dot{D}}~.
\end{split}
\end{equation}

A fully antisymmetric Levi-Civita symbol \(\ve^{abcd}\) is normalized by 
\(\ve^{1234}=1\). The (anti-)self-dual parts of an antisymmetric tensor 
$T_{ab}$ 
are~\footnote{Our sign 
convention here is opposite to that in refs.~\cite{rev1} and \cite{hat}.}
\begin{equation}
	\Tilde{T}{}_{ab}=\frac{1}{2}\ve_{abcd}T^{cd}~,
\qquad
	T^{\pm}_{ab}=\frac{1}{2}(T_{ab}\pm \Tilde{T}{}_{ab})
\end{equation}

\section{Consistent freezing of a graviphoton field}

Our basic idea is to eliminate a graviphoton field from the $N=2$ 
matter-coupled supergravity, by assigning it a fixed value, say, its vacuum 
expectation value (VEV). Generally speaking, it is  going to
break supersymmetry, because the graviphoton is a field component of $N=2$ 
supergravity multiplet. We show in this section that, when assigining a 
self-dual vacuum expectation value to a graviphoton, the $N=2$ local 
supersymmetry can be consistently broken to an $N=1/2$ local supersymmetry in 
four Euclidean dimensions, and in the presence of a generic $N=2$ matter, 
thus generalizing the earlier 
results \cite{hat} obtained for pure $N=2$ supergravity without matter.

The graviphoton field is identified with $W_\m^0(x)$ after imposing the gauge 
condition on 
$X^0$ by using the local chiral $U(1)$ rotations \cite{hol}: 
\be
X^0=\Bar{X}{}^0 > 0 \label{grgauge}
\ee
We prefer to impose the gauge condition (\ref{grgauge}) {\it after} our 
truncation procedure described in this section. The graviphoton field strength 
is given by 
\be
	F_{\m\n}^0=\pa_\m W^0_\n-\pa_\n W^0_\m 
\ee

We now freeze the graviphoton field by imposing the self-dual constraints: 
\be
\label{C-defo}
	F_{\m\n}^{0+}\equiv C_{\m\n}(x) 
\qquad{\rm and}\qquad
	F_{\m\n}^{0-}\equiv 0~~, 
\ee
where $C_{\m\n}(x)$ is a fixed self-dual antisymmetric tensor field (VEV) 
with an arbitrary $x$-dependence.

Our strategy is to look for the {\it residual} local supersymmetry that keeps
 our assignment (\ref{C-defo}) 
invariant, i.e. that leaves both $F_{\m\n}^{0+}$ and $F_{\m\n}^{0-}$ to be 
unchanged. As was shown in 
ref.~\cite{hat} in the case of the $N=2$ supergravity wihout matter, we already
 have to fix $3/4$ of 
$Q$-supersymmetry by eliminating three (out of four) infinitesimal local
 $Q$-supersymmetry chiral spinor 
parameters as follows:
\be
	\ve^1 \equiv \ve^2 \equiv \Bar{\ve}{}_2 \equiv 0
\ee
Then we may only hope that the remaining local $N=1/2$ supersymmetry with the
 chiral spinor parameter 
$\Bar{\ve}{}_1(x)$ remains to be a symmetry of the theory after investigating 
all the consistency
conditions originiating from the $C$-deformation (\ref{C-defo}):
\be
\d_{\Bar{\ve}{}_1} F^0_{\m\n}=0
\ee

The $N=1/2$ $Q$-supersymmetry transformation law of $W_{\m}^0$ can be read off 
from eq.~(\ref{app:sc.W}):
\be
	\d_{\Bar{\ve}{}_1} W^0_\m
	=
	-i\Bar{\ve}{}_1\s_\m\O_2^0-2\Bar{\ve}{}_1\Bar{\j}{}_{\m 2}X^0
\ee
It implies the following consistency condition:
\be
	(\s_\m\O_2^0){}_{\dot{A}}-2i\Bar{\j}{}_{\m 2\dot{A}}X^0=0
\label{defocon1}
\ee
This algebraic condition can be easily solved, thus eliminating an independent 
gravitino field 
$\Bar{\j}{}_{\m 2\dot{A}}$, in terms of the other (matter) fields.

Equation (\ref{defocon1}) is also not invariant under the $N=1/2$ 
$Q$-supersymmetry, so that
we have to impose its invariance for consistency. By using the $N=1/2$
 $Q$-supersymmetry transformation 
laws 
\be
	\d_{\Bar{\ve}{}_1} e_\m^m
	=
	-i\Bar{\ve}{}_1\s^m\j_\m^1~,
\quad
	\d_{\Bar{\ve}{}_1} \O_2^{0 A}=0~,
\quad
	\d_{\Bar{\ve}{}_1} \Bar{\j}{}_{\m 2\dot{A}}
	=
	\cv_{\m 2}{}^1\Bar{\ve}{}_{1\dot{A}}~,
\quad
	\d_{\Bar{\ve}{}_1} X^0=0~, 
\ee
we find from eqs.~(\ref{app:sc.e}), (\ref{app:sc.Bar{j}_2}), (\ref{app:sc.X})
 and (\ref{app:sc.O_2})
that the constraint (\ref{defocon1}) yields yet another constraint 
\be
	\j_\m^1\O_2^0-\cv{}_{\m 2}{}^1 X^0=0
\label{defocon2}
\ee
This is again an algebraic equation, while it can be easily solved for
 $\cv{}_{\m 2}{}^1$, thus eliminating
that field in terms of the other fields.

Generally speaking, further consistency requirements might lead to the
 infinite and increasingly 
complicated set of constraints, but it does not happen in our case! We find
 that 
eq.~(\ref{defocon2}) is invariant under the $N=1/2$ $Q$-supersymmetry because
 of
\be
	\d_{\Bar{\ve}{}_1} \j_\m^1=0
\quad {\rm and}\quad
	\d_{\Bar{\ve}{}_1} \cv_{\m 2}{}^1=0
\ee
which easily follow from eqs.~(\ref{app:sc.j^1}) and (\ref{app:last}),
 respectively.

Having found the consistent short (finite) set of algebraic constraints that
 are invariant under the
$N=1/2$ local $Q$-supersymmetry, we can easily check what are the other
 residual symmetries in the
list of the $N=2$ superconformal transformation laws (see Appendices B and C),
 which also leave the 
constraints invariant.  We then find that all our constraints (\ref{C-defo}),
 (\ref{defocon1}) and 
(\ref{defocon2}) are still invariant under the full $S$-supersymmetry and the
 chiral $U(1)$ 
transformations, whereas the local $SU(2)$ automorphisms of the $N=2$ 
supersymmetry algebra are broken: $\L^i{}_j=\L_i{}^j=0$.

The residual $S$-supersymmetry implies that the original decomposition laws 
(see ref.~\cite{hol} and 
Appendix C) are still valid, being subject to the conditions
 $\ve^1=\ve^2=\Bar{\ve}{}_2=0$. It means 
that our construction does not affect the $S$-gauge, $K$-gauge and $D$-gauge,
 as described in
ref.~\cite{hol}. So we are going to use the same gauges when passing to the
 Poincar\'e supergravity.

The rest of our construction of the $N=1/2$ matter-coupled supergravity is
 pretty straightforward
(though tedious!), by inserting the equations derived in this section into
 the results of ref.~\cite{hol},
making gauge-fixing, and deriving the transformation rules and the Lagrangian
 of the (Poincar\'e)
$N=1/2$ supergravity with matter. Our results are summarized in the next
 Secs.~5, 6 and 7.

\section{$N=1/2$ supergravity transformation laws}

Here we summarize the residual local superconformal symmetry transformation
 laws of the independent 
field components with respect to the $Q$-supersymmetry with the parameters 
$(\Bar{\ve}_1)$, 
the $S$-supersymmetry with the parameters $(\eta_i)$ and $(\Bar{\eta}^i)$, 
and the chiral $U(1)$ symmetry 
with the parameter $(\L_A)$.

\noindent (i) As regards a vierbein and gravitini, we find
\begin{alignat*}{3}
&\d e_\m{}^a
&=&
	-i\Bar{\ve}{}_1\s^a\j_\m^1~,	\\
&\d \j_\m^{1A}
&=&
	+i(\s_\m\Bar{\h}{}^1)^A
	-\frac{1}{2}i\L_A\j^{1A}_\m~,	\\
&\d \j_\m^{2A}
&=&
	-\frac{1}{2}iT^{-21}_{\m m}(\s^m\Bar{\ve}{}_1)^A
	+i(\s_\m\Bar{\h}{}^2)^A
	-\frac{1}{2}i\L_A\j^{2A}_\m,		\\
&\d \Bar{\j}{}_{\m 1\dot{A}}
&=&
	+2\bigl[
		\pa_\m\Bar{\ve}{}_{1\dot{A}}
		-
		\frac{1}{2}\o_\m^{mn}\s_{mn}{}_{\dot{A}}{}^{\dot{B}}
		\Bar{\ve}{}_{1\dot{B}}
		-
		\frac{1}{2}i\Bar{A}{}_\m\Bar{\ve}{}_{1\dot{A}}
	\bigr]
	+\cv{}_\m{}_1{}^1\Bar{\ve}_{1\dot{A}}
\\
& & &
	-i(\s_\m\h_1)_{\dot{A}}
	+\frac{1}{2}i\L_A\Bar{\j}{}_{\m 1\dot{A}}
\end{alignat*}
As is clear from those equations, the gravitino field 
 $\Bar{\j}{}_{\m 1\dot{A}}$
is the gauge field of the residual $N=1/2$ local supersymmetry.	

\noindent (ii) As regards the vector multiplet components, we find $(I>0)$ 
\begin{alignat*}{2}
&	\d X^I
	&=&
	-i\L_A X^I~,			\\
&	\d \Bar{X}{}^I
	&=&
	+\Bar{\ve}{}_1\Bar{\O}{}^{1I}
	+i\L_A \Bar{X}{}^I~,		\\
&	\d W_\m^I
	&=&
	-i \Bar{\ve}{}_1 \s_\m \O_2^I  
	-2\Bar{\ve}{}_1\Bar{\j}_{\m 2}X^I~,	\\
&	\d \O_1^{IA}
	&=&
	-2i(\s^\l\Bar{\ve}{}_1)^A
	[
		(\pa_\l+iA_\l)X^I
		-
		gf_{JK}{}^I W_\l^J X^K 
		-
		\frac{1}{2} \j^k_\l\O^I_k
	]
\\
& & &	+2X^I\h_1^A
	-\frac{1}{2}i\L_A \O_1^{IA}~,	\\
&	\d \O_2^{IA}
	&=&
	+2X^I\h_2^A
	-\frac{1}{2}i\L_A \O_2^{IA}~,	\\
&	\d \Bar{\O}{}^{1I \dot{A}}
	&=&
	+
	Y{}^{11I}\Bar{\ve}{}_1^{\dot{A}}
	+
	2\Bar{X}{}^I \Bar{\h}{}^{1\dot{A}}
	+
	\frac{1}{2}i\L_A \Bar{\O}{}^{1I \dot{A}}~,	\\
&	\d \Bar{\O}{}^{2I \dot{A}}
	&=&
	+
	Y{}^{21I}\Bar{\ve}{}_1^{\dot{A}}
	+
	\Bar{\ve}{}_1^{\dot{B}}\s^{mn}{}_{\dot{B}}{}^{\dot{A}}\cf^{+I}_{mn}
	+
	2gf_{JK}{}^I \Bar{X}{}^J X^K\Bar{\ve}{}_1^{\dot{A}}
\\
& & &	+
	2\Bar{X}{}^I \Bar{\h}{}^{2\dot{A}}
	+
	\frac{1}{2}i\L_A \Bar{\O}{}^{2I \dot{A}}
\end{alignat*}

\noindent (iii) As regards the hypermuliplet components, we find 
(see Appendix A for the notation)
\begin{alignat*}{2}
&	\d A_1{}^\a
	&=&
	-2\Bar{\z}{}^\a\Bar{\ve}{}_1~,	\\
&	\d A_2{}^\a
	&=&
	0~,				\\
&	\d A^1{}_\a
	&=&
	0~,				\\
&	\d A^2{}_\a
	&=&
	+2\r_{\a\b}\Bar{\ve}{}_1\Bar{\z}{}^\b~,	\\
&	\d\z_{\a A}
	&=&
	+i(\Bar{\ve}{}_1\s^\m)_{A}\Bar{D}{}_\m A{}^1_\a
	+
	A^i{}_\a\h_{iA}
	+
	\frac{i}{2}\L_A \z_{\a A}~,		\\
&	\d \Bar{\z}{}^\a_{\dot{A}}
	&=&
	-
	2gX^\a{}_\b A_2{}^\b \Bar{\ve}{}_{1\dot{A}}
	+
	A_i{}^\a\Bar{\h}{}^i_{\dot{A}}
	-
	\frac{i}{2}\L_A \Bar{\z}{}^\a_{\dot{A}}
\end{alignat*}
where we have used the abbreviations (B.13), (B.14) and (B.15).

\section{Gauge-fixing and eliminating auxiliary fields}

In the superconformal tensor calculus, the $N=1$ or $N=2$ matter-coupled 
(Poincar\'e) supergravity is 
obtained from the $N=1$ or $N=2$ conformal supergravity, respectively,
 by imposing certain gauges, in
order to fix the truly superconformal symmetries, while keeping the
 super-Poincar\'e symmetries. In this
process the residual supersymmetry transformations are deformed by the
 compensating transformations
needed to restore the gauges \cite{hol}. We follow here the same pattern by
 eliminating the truly 
superconformal $N=1/2$ symmetries after imposing our set of constraints
 (Sec.~4), with the help of the 
gauges similar to that of ref.~\cite{hol}. Then we eliminate the auxiliary
 fields by using their algebraic equations of motion. The Lagrangian of the 
resulting $N=1/2$ supergravity is given in the 
next section. 

\begin{itemize}
	\item The $K$-gauge to fix the conformal boosts is given by
	\be b_\m=0 \ee
	\item The $D$-gauge to fix dilatations (and to get the standard 
normalization of the Einstein term
in the Lagrangian) is given by
	\be	
		N_{IJ}X^I\Bar{X}^J=1, 
	 	\quad A_i{}^\a d_\a{}^\b A^i{}_\b=-2
	\ee
	\item The $S$-gauge to fix $S$-supersymmetry is
	\begin{align}
		&	X^I N_{IJ}\Bar{\O}{}^{iJ}_{\dot{A}}=0~, 
		&	\Bar{X}{}^I N_{IJ}\O_i^{J A}=0~, 
		&\quad	A^i{}_\a d^\a{}_\b \Bar{\z}{}^\b_{\dot{A}}=0~, 
		&	\z_\b^A d_\a{}^\b A_i{}^\a=0
	\end{align}
\end{itemize}
where we have used the notation \cite{hol}
	\(N_{IJ}=\frac{1}{4}(F_{IJ}+\Bar{F}_{IJ})\)
in terms of a homogenous function \(F(X^I)\) of degree two in $X^I$. 
Here the subscripts $I, J, \dots$ stand for
the derivatives with respect to $X^I, X^J, \dots$, respectively. The function
 \(F(X^I)\) obeys the relations 
\begin{alignat*}{2}
 	F(X)&=\half F_I(X)X^I~,
&	F_I(X)&=F_{IJ}(X)X^J~,
\\
 	F_{IJK}(X)X^K&=0~,
&	F_{IJK}(X)&=-F_{IJKL}(X)X^L
\end{alignat*}

As a result, the $S$-supersymmetry parameters can be written down in terms of
the $N=1/2$ $Q$-supersymmetry parameter  \(\Bar{\ve}{}_1\) as follows:
\begin{alignat*}{3}
&	\h_{1 A}
	&=&
	-(\Bar{\ve}{}_1\s^\m)_A
	\Bigl[
		\frac{1}{8}N_{IJ}(
			\O_1^I\s_\m\Bar{\O}{}^{1I}
			-
			\O_2^I\s_\m\Bar{\O}{}^{2J}
			)
		+d^\a{}_\b(\z_\a\s_\m\Bar{\z}{}^\b)
	\Bigr]~,						\\
&	\h_{2 A}
	&=&
	-\frac{1}{4}(\Bar{\ve}{}_1\s^\m)_A 
	N_{IJ}\O_2^I\s_\m\Bar{\O}{}^{1J}~,		\\
&	\Bar{\h}{}^1_{\dot{A}}
	&=&
	+2gd^\a{}_\b\Bar{\ve}{}_{1\dot{A}}
	A_2{}^\a X_\b{}^\g A^1{}_\g~,			\\
&	\Bar{\h}{}^2_{\dot{A}}
	&=&
	-2gd^\a_\b\Bar{\ve}{}_{1\dot{A}}
	A_1{}^\a X_\b{}^\g A^1{}_\g
	-d_\a{}^\b\r_{\b\g}
	(\s^{\r\l}\Bar{\ve}{}_1)_{\dot{A}}
	(\Bar{\z}^{\a}\s_{\r\l}\Bar{\z}^{\g})
\end{alignat*}

Similarly, as regards the chiral $U(1)$ rotations, we find
\begin{equation*}
	\L_A=\frac{i}{2}N_{IJ}\Bar{\ve}{}_1 \Bar{\L}{}^{1I}X^J
\end{equation*}

The important part of the superconformal tensor calculus is a construction of
the (superconformally) invariant actions. As regards the $N=2$ case, the 
invariant
action for vector multiplets is given by eq.~(3.9) of ref.~\cite{hol}, whereas
the invariant action of hypermultiplets is given by eq.~(3.29) of 
ref.~\cite{hol}.
The full invariant action for vector- and hyper-multiplets coupled to 
$N=2$ conformal
supergravity is a sum of eq.~(3.9), a Chern-Simons coupling (3.16), and 
eq.~(3.29) of
ref.~\cite{hol}. We use the same $N=2$ superconformally invariant action as 
our starting 
point. Hence, we are still in a position to fix the algebraic field equations 
of the 
auxiliary fields that follow
from the full action \cite{hol} after taking into account our constraint 
(Sec.~4).~\footnote{It is always assumed here that we are working in Euclidean 
space. 
Hence, the results of ref.~\cite{hol} are to be reformulated in the Euclidean 
signature with the 2-component notation for spinors -- see Sec.~3.}

As regards the chiral $U(1)$ gauge fields, we find 
\begin{equation}
	A_\m
	=
	\frac{i}{2}N_{IJ}[X{}^I\hat{\pa}{}_\m 
\Bar{X}{}^J-(\hat{\pa}{}_\m X{}^I)\Bar{X}{}^J]
	-
	\frac{1}{8}N_{IJ}\Bar{\O}{}^{iI}\s_\m\O^J_i
	-
	d_\a{}^\b\Bar{\z}{}^\a\s_\m\z_\b
\end{equation}

As regards the chiral $SU(2)$ gauge fields,  the $\cv_{\m 2}{}^1$ is already 
fixed (Sec.~4) by the
constraint (\ref{defocon2}) whose solution is 
\begin{equation}
	\cv_{\m 2}{}^1=\fracmm{1}{X^0}(\j_\m^1\O_2^0)
\end{equation}
The remaining  $SU(2)$ gauge fields are given by (when $i\neq 2$ and $j\neq 1$)
 
\begin{equation}	
	\cv_\m{}_i{}^j
	=
	d^\b{}_\a (\pa_\m A^j{}_\b A_i{}^\a - A^j{}_\b\pa_\m A_i{}^\a)
	-
	\frac{i}{2}N_{IJ}\Bar{\O}{}^{jI}\s_\m \O_i{}^J
	+
	\frac{i}{4}\d_i^j N_{IJ}\Bar{\O}{}^{kI}\s_\m \O_k^J
\end{equation}

The $T^{+ij}_{\m\n}$ after the $C$-deformation ($F_{\m\n}^{0+}= C_{\m\n}$) 
is determined
by the algebaric equation
\begin{equation}
\begin{split}
	N_{IJ}X^I X^J T^{+}_{\m\n}{}^{ij}\ve_{ij}
&	-4N_{I0}X^I C_{\m\n}
	-4N_{I\Tilde{J}}X^I \hat{F}{}^{+}_{\m\n}{}^{\Tilde{J}}
\\
&	-8d_\a{}^\b\r_{\b\g}\Bar{\z}{}^\a\s_{\m\n}\Bar{\z}{}^\g
	-\frac{1}{4}\Bar{F}{}_{IJK}X^K\ve_{ij}
	\Bar{\O}{}^{iI}\s_{\m\n}\Bar{\O}{}^{jJ}
	=0~,
\end{split}
\end{equation}
whereas for the $T^{-}_{\m\n ij}$ after the $C$-deformation 
($F_{\m\n}^{0-}= 0$) we find 
\begin{equation}
\begin{split}
	N_{IJ}\Bar{X}{}^I \Bar{X}{}^J T^{-}_{\m\n ij}\ve^{ij}
&	-4N_{I\Tilde{J}}\Bar{X}{}^I \hat{F}{}^{-}_{\m\n}{}^{\Tilde{J}}
\\
&	+8d^\a{}_\b\r^{\b\g}\z_\g\s_{\m\n}\z_\a
	+\frac{1}{4}F_{IJK}\Bar{X}{}^K\ve^{ij}
	\O{}_i^I\s_{\m\n}\O{}^J_j
	=0
\end{split}
\end{equation}

Finally, the vector multiplet auxiliary fields obey the algebraic field 
equations
\begin{equation}
	-\frac{1}{4}N_{IJ}Y_{ij}^J
	+
	g d_\a{}^\b A^k{}_\b\ve_{ki}t_I{}^\a{}_\g A_j{}^\g
	-
	\frac{1}{32}(
		F_{IJK}\O_i^J\O_j^K 
		+ 
		\Bar{F}_{IJK}\ve_{ik}\ve_{jl}\Bar{\O}{}^{kJ}\Bar{\O}{}^{lK}
	)
	=0~,
\end{equation}
and
\begin{equation}
	-\frac{1}{4}N_{IJ}Y^{ijJ}
	-
	g d^\a{}_\b A^j{}_\g\ve^{ik}t_I{}_\a{}^\g A_k{}^\b
	+
	\frac{1}{32}(
		\Bar{F}{}_{IJK}\Bar{\O}{}_i^J\Bar{\O}{}_j^K 
		+ 
		\Bar{F}_{IJK}\ve^{ik}\ve^{jl}\O{}_k^J\O{}_l^K
	)
	=0
\end{equation}

\section{Lagrangian}

A derivation of the full Lagrangian of the new $N=1/2$ supergravity with 
vector- and hyper-multiplets is now fully straightforward, so we merely 
present our final result.

Let $\hat{\pa}{}_\m$ be the covariant derivative with respect to the 
non-abelian gauge transformations and local Lorentz rotations, but not w.r.t. 
the chiral $U(1)$ and $SU(2)$ rotations. The gravitino field  
$\Bar{\j}{}_{\m 2}$ is not independent, but a solution to eq.~(\ref{defocon1}).
In our Lagrangian $\Bar{\j}{}_{\m 2}$ is just the notation for  
$\Bar{\j}{}_{\m 2 \dot{A}}=-\fracmm{i}{2X^0}(\s_\m\O_2^0){}_{\dot{A}}$.

The $N=1/2$ supergravity Lagrangian has the following structure:
\be
	\cl=	\cl{}_{\rm kin}+\cl{}_{\rm 4-fermi}+\cl{}_{\rm contact}
		+\cl{}_{\rm gauge}+\cl{}_{\rm F}+\cl{}_{\rm aux} 
\label{flag}~~,
\ee
whose separate terms read as follows: 
\begin{align*}
	e^{-1} \cl{}_{\rm kin}
	=
&	-\frac{1}{2}R 
	-d_\a{}^\b\hat{\pa}{}_\m A^i{}_\b\hat{\pa}{}^\m A_i{}^\a
	+ N_{IJ}\hat{\pa}{}_\m X^I\hat{\pa}{}^\m \Bar{X}{}^J
\\		
&	+ie^{-1}\ve^{\m\n\r\s}
	\j_\m^i\s_\n\hat{\pa}{}_\r \Bar{\j}{}_{\s i}	\\
&	-\frac{i}{4}N_{IJ}\Bar{\O}{}^{iI}\s^\m\hat{\pa}{}_\m\O_i^J
	+\frac{i}{4}N_{IJ}\O_i^I\s^m\hat{\pa}{}_\m\Bar{\O}{}^{iJ} \\
&	+2id_\a{}^\b\Bar{\z}{}^\a\s^\m\hat{\pa}{}_\m\z_\b
	+2id^\a{}_\b(\hat{\pa}{}_\m\Bar{\z}{}^\b)\s^\m\z_\a~,
\end{align*}
\begin{align*}
	e^{-1} \cl{}_{\rm 4-fermi}
	=
&	+\frac{1}{16}N_{IJ}e^{-1}\ve^{\m\n\r\s}
	(\Bar{\j}{}_{\m i}\s_\n\O_j\ve^{ij})
	(\Bar{\j}{}_{\r k}\s_\s\O_l^J\ve^{kl}) 		\\
&	+\frac{1}{16}N_{IJ}e^{-1}\ve^{\m\n\r\s}
	(\Bar{\O}{}^{jI}\s_\n\j_\m^i\ve_{ij})
	(\Bar{\O}{}^{lK}\s_\s\j_\r^k\ve_{kl})		\\
&	+\frac{1}{8}N_{IJ}e^{-1}\ve^{\m\n\r\s}
	(\Bar{\j}{}_{\m i}\Bar{\j}{}_{\n j}\ve^{ij})
	\Bigl[
		-i\Bar{\j}{}_{\r k}\s_\s\O_l^I X^J
		-\frac{1}{2}\Bar{\j}{}_{\r k}\Bar{\j}{}_{\s l}X^I X^J
	\Bigr]\ve^{kl}					\\
&	-\frac{1}{8}N_{IJ}e^{-1}\ve^{\m\n\r\s}
	(\j_\m^i \j_\n^j\ve_{ij})
	\Bigl[
		-i\Bar{\O}{}^{lI}\s_\s\j^k_\r\Bar{X}{}^J
		+\frac{1}{2}\j_\r^k \j_\s^l\ve_{kl}
		\Bar{X}{}^I\Bar{X}{}^J
	\Bigr]\ve_{kl}					\\
&	+\frac{i}{48}F_{IJK}
	(\Bar{\j}{}_{i\m}\s^\m\O_k^I)(\O^J_l\O^K_j)
	\ve^{ij}\ve^{kl}				\\
&	-\frac{i}{48}\Bar{F}{}_{IJK}
	(\Bar{\O}{}^{kI}\s^\m\j_\m^i)
	(\Bar{\O}{}^{jK}\Bar{\O}{}^{lJ})
	\ve_{ij}\ve_{kl}				\\
&	-d_\a{}^\b(\Bar{\z}{}^\a\s^\m\s^\n\Bar{\j}{}_{\m i})
	(
	\z_\b\j^i_\n
	+
	\r_{\b\g}\Bar{\z}{}^\g\Bar{\j}{}_{\n j}\ve^{ij}
	)						\\
&	+d^\a{}_\b(\j_\m^i\s^\n\s^\m\z_\a)
	(
	\Bar{\j}{}_{\n i}\Bar{\z}{}^\b
	+
	\r^{\b\g}\j_\n^j\z_\g\ve_{ij}
	)						\\
&	-\frac{1}{192}F_{IJKL}(\O_i^I\O_k^J)(\O_j^K\O_l^L)
	\ve^{ij}\ve^{kl}				\\
&	-\frac{1}{192}\Bar{F}{}_{IJKL}
	(\Bar{\O}{}^{iI}\Bar{\O}{}^{kJ})
	(\Bar{\O}{}^{jK}\Bar{\O}{}^{lL})\ve_{ij}\ve_{kl}~,
\end{align*}
\begin{align*}
	e^{-1}\cl{}_{\rm contact}
	=
&	+\frac{i}{4}e^{-1}\ve^{\m\n\r\s}\Bar{\j}{}_{\m i}\s_\n\j_\r^i 
	N_{IJ}[X^I(\hat{\pa}{}_\s \Bar{X}{}^J)
-(\hat{\pa}{}_\s X^I)\Bar{X}{}^J]	\\
&	+\frac{i}{2}e^{-1}\ve^{\m\n\r\s}\Bar{\j}{}_{\m i}\s_\n\j_\s^jd_\a{}^\b
	(A_j{}^\a \hat{\pa}{}_\s A^i{}_\b-A^i{}_\b\hat{\pa}{}_\s A_j{}^\a)
		\\
&	-\frac{1}{2}N_{IJ}\j_\m^i\s^\n\s^\m\O_i^J\hat{\pa}{}_\n \Bar{X}{}^I
-\frac{1}{2}N_{IJ}\Bar{\O}{}^{iJ}\s^\m\s^\n\Bar{\j}{}_{\m i}\hat{\pa}{}_\n X^I
	\\
&	+2d_\a{}^\b\j^i_\m\s^\n\s^\m\z_\b\hat{\pa}{}_\n A_i{}^\a
	+2d_\a{}^\b\Bar{\z}{}^\b\s^\m\s^\n\Bar{\j}{}_{\m i}
\hat{\pa}{}_\n A^i{}_\a		\\
&	+\frac{1}{8}N_{IJ}(\Bar{\O}{}^{iI}\s^\m\s^\n\Bar{\j}{}_{\m i})
(\j^j_\n\O_j^J)
	-\frac{1}{8}N_{IJ}(\j^i_\m\s^\n\s^\m\O_i^I)(\Bar{\O}{}^{jJ}
\Bar{\j}{}_{\n j})		\\
&	+\frac{i}{16}F_{IJK}\O_i^I\s^\m\Bar{\O}{}^{iK}\hat{\pa}{}_\m 
X^J	
	+\frac{i}{16}\Bar{F}{}_{IJK}\O_i^K\s^\m\Bar{\O}{}^{iI}\hat{\pa}{}_\m 
\Bar{X}{}^J~,
\end{align*}
\begin{align*}
	e^{-1}\cl_{\rm gauge}
	=
&	-\frac{i}{6}gC_{I,JK}e^{-1}\ve^{\m\n\r\s}W^I_\m W^J_\n
	\Bigl(
		\pa_\r W_\s^K
		-\frac{3}{8}gf_{LM}{}^K W_\r^L W_\s^M
	\Bigr)						\\
&	+4g^2d_\a{}^\b A^i{}_\b \Bar{X}{}^\a{}_\g X^\g_\d A_i{}^\d
	-g^2N_{IJ}f_{KL}{}^I\Bar{X}{}^K X^L f_{MN}{}^J \Bar{X}{}^M X^N
		\\
&	-\frac{1}{2}gN_{IJ}\O_i^I f_{KL}{}^J\Bar{X}{}^K\O_j^L\ve^{ij}
	+\frac{1}{2}gN_{IJ}\Bar{\O}{}^{jL}X^Kf_{KL}{}^J\Bar{\O}{}^{iI}\ve_{ij}
	\\
&	+4gd_\a{}^\b A^i{}_\b \Bar{\O}{}^{j\a}{}_\g \Bar{\z}{}^\g\ve_{ij}
	-4gd^\a{}_\b A_i{}^\b\z_\g\O_{j\a}{}^\g\ve^{ij}	\\
&	-4gd_\a{}^\b\r_{\b\g}\Bar{\z}{}^\a\Bar{X}{}^\g{}_\d\Bar{\z}{}^\d
	+4gd^\a{}_\b\r^{\b\g}\z_\d X_\g{}^\d\z_\a	\\
&	-\frac{i}{2}gN_{IJ}\j^i_\m\s^\m\Bar{\O}{}^{jI}\ve_{ij}f_{KL}{}^J X^K
\Bar{X}{}^L
	-\frac{i}{2}gN_{IJ}\O_j^I\s^\m\Bar{\j}{}_{\m i}\ve^{ij}f_{KL}{}^J 
\Bar{X}{}^K X^L	\\
&	+4igd_\a{}^\b\j^i_\m\s^\m\Bar{\z}{}^\g A^j{}_\b\Bar{X}{}^\a{}_\g
\ve_{ij}
	+4igd^\a{}_\b\z_\g\s^\m\Bar{\j}{}_{\m i} X_\a{}^\g A_j{}^\b\ve^{ij}
		\\
&	+igd_\a{}^\b\j_\m^i\s^\m\Bar{\O}{}^{k\a}{}_\g A^j{}_\b A_k{}^\g\ve_{ij}
	+igd^\a{}_\b\O_{k\a}{}^\g\s^\m\Bar{\j}{}_{\m i}A^k{}_\g A_j{}^\b
\ve^{ij}	\\
&	+2gd_\a{}^\b\j^i_\m\s^{\m\n}\j_\n^j A_i{}^\a A^k{}_\g 
\Bar{X}{}_\b{}^\g\ve_{jk}
	-2gd^\a{}_\b\Bar{\j}{}_{\n j}\s^{\m\n}\Bar{\j}{}_{\m i}X^\b{}_\g 
A_k{}^\g A^i{}_\a\ve^{jk}~,
\end{align*}
where \(C_{I,JK}\) are the real coefficient functions defined in terms of the 
input function $F(X)$ 
by considering the non-abelian gauge transformations of the latter 
(with the gauge coupling constant $g$) \cite{hol}:
\begin{equation}
\d F =gF_J f_{IK}{}^J X^K\L^I \equiv ig\L^I C_{I,JK}X^JX^K \label{C-coeff}
\end{equation}
The coefficient functions  \(C_{I,JK}\) are symmetric in the last two indices,
 and obey a relation \cite{hol}
\begin{equation}
	C_{I,JK}+C_{J,KL}+C_{K,IJ}=0
\end{equation}

The $\cl{}_{\rm F}$ part of the Lagrangian (\ref{flag}) is given by
\begin{align*}
	e^{-1}\cl{}_{\rm F}
	=
&	-\frac{1}{4}
	\Bigl[
	N_{I0}C^{\m\n}
	+N_{I\Tilde{J}}\Tilde{F}{}^{\m\n\Tilde{J}}
	N_{IJ}\Tilde{G}{}^{\m\n J}
	\Bigr]
	G_{\m\n}^I				\\
&	+\frac{1}{8}
	\Bigl[
		N_{00}(C^{\m\n}+G^{\m\n 0})
		+N_{0\Tilde{J}}\hat{F}{}^{\m\n\Tilde{J}}
	\Bigr]
	(C_{\m\n}+G_{\m\n}^0)		\\
&	+\frac{1}{8}
	\Bigl[
		N_{\Tilde{I}0}(C^{\m\n}+G^{\m\n 0})
		+N_{\Tilde{I}\Tilde{J}}\hat{F}{}^{\m\n\Tilde{J}}
	\Bigr]
	\hat{F}{}_{\m\n}^{\Tilde{I}}		\\
&	+\frac{1}{32}
	\Bigl[
		\Bar{F}{}_{IJ0}C_{\m\n}
		+\Bar{F}{}_{IJ\Tilde{K}}F^{+\Tilde{K}}_{\m\n}
		+\Bar{F}{}_{IJK}G^{+K}_{\m\n}
	\Bigr]
	(\Bar{\O}{}^{iI}\s^{\m\n}\Bar{\O}{}^{jJ}\ve_{ij})	\\
&	-\frac{1}{32}
	\Bigl[
		F_{IJ\Tilde{K}}F^{-\Tilde{K}}
		+F_{IJK}G^{-K}_{\m\n}
	\Bigr]
	(\O_j^J\s^{\m\n}\O_i^I\ve^{ij})			\\
&	-\frac{1}{32}(F_{00}-\Bar{F}{}_{00})C_{\m\n}C^{\m\n}
	-\frac{1}{16}(F_{\Tilde{I}0}-\Bar{F}{}_{\Tilde{I}0})
	F^{\Tilde{I}}_{\m\n}C^{\m\n}			\\
&	-\frac{1}{64}(F_{\Tilde{I}\Tilde{J}}-\Bar{F}{}_{\Tilde{I}\Tilde{J}})
	e^{-1}\ve^{\m\n\r\s}F^{\Tilde{I}}_{\m\n}F^{\Tilde{J}}_{\r\s}
\end{align*}
where we have introduced the non-abelian vector field strength
\be
	F^{\Tilde{I}}_{\m\n}
	=
	\pa_\m W_\n^{\Tilde{I}}-\pa_\n W_\m^{\Tilde{I}}
	-gf_{\Tilde{J}\Tilde{K}}{}^{\Tilde{I}}
	W_\m^{\Tilde{J}}W_\n^{\Tilde{K}}
\ee
and the book-keeping notation
\be
	G_{\m\n}^I
	:=
	-\frac{i}{2}
	\Bigl[
		\Bigl(
			\Bar{\O}{}^{iI}\s_\m\j_\n^j\ve_{ij}
			-\O_i^I\s_\m\Bar{\j}{}_{\n j}\ve^{ij}
			-(\m \leftrightarrow \n)
		\Bigr)
	\Bigr]
	+
	\Bigl(
		\Bar{X}{}^I\j^i_\m\j^j_\n\ve_{ij}
		-X^I\Bar{\j}{}_{\m i}\Bar{\j}{}_{\n j}\ve^{ij}
	\Bigr)
\ee

Finally, the last term $\cl_{\rm aux}$ in eq.~(\ref{flag}) is given by
\begin{align*}
	e^{-1}\cl_{\rm aux}
	=
&	-A_\m A^\m
	-\frac{1}{4}\cv_\m{}^i{}_j\cv^\m{}_j{}^i
	+\frac{1}{8}N_{IJ}Y^I_{ij}Y^{ijJ}	\\
&	-\frac{1}{64}N_{IJ}X^I X^J (T^{+}_{\m\n ij}\ve^{ij})^2
	-\frac{1}{64}N_{IJ}\Bar{X}{}^I \Bar{X}{}^J (T^{-ij}_{\m\n}\ve_{ij})^2
\end{align*}

The gravitino field $\j_\m^2$ enters the Lagrangian (\ref{flag}) algebraically,
 so it may be eliminated via its non-propagating field equation. The bosonic 
part of the matrix multiplying  $\j_\m^2$ in its field equation has an inverse,
 due to the identity
\be
	\bigl(
		2\s_{\r\l}{}^C{}_A
		-g_{\r\l}\d^C{}_A
		-\frac{1}{3}\s_\r^{C\dot{D}}\s_{\l\dot{D}A}
	\bigr)
	\s^{\l\m A}{}_B
	=
	+\d_\r^\m \d^C{}_B
\ee
so that there is a unique solution for  $\j_\m^2$. 

\newpage

\section{Conclusion}

We formulated the $N=1/2$ supergravity coupled to both vector and scalar
matter multiplets in four Euclidean dimensions. The gauge field of the local 
$N=(0,\frac{1}{2})$ supersymmetry is given by a single chiral gravitino. 
The $C$-deformation, originally introduced in the context of supersymmetric
D-branes with RR-type fluxes, was the main tool of our purely field-theoretical
 construction. The new matter-coupled $N=1/2$ supergravity with matter is not 
invariant under the Euclidean analogue of Lorentz rotations in four dimensions,
 due to the explicit presence of a fixed (self-dual) graviphoton background. In
 addition, the Lagrangian we constructed (Sec.~7) is not Hermitean, thus 
hampering immediate physical applications of the proposed new supergravity 
theory. However, those are the common problems of all recently constructed 
$N=1/2$ supersymmetric field theories, either with rigid or local $N=1/2$ 
supersymmetry.

When compared to our earlier construction of the pure $N=1/2$ supergravity
without matter \cite{hat}, in the matter-coupled $N=1/2$ supergravity we 
observe no formation of the gravitino condensate. Instead, we got one more
constraint on the gravitini in terms of the matter fields.
 
Being the first construction of that type, our $N=1/2$ supergravity with matter
is unlikely to be the most general one having a local $N=1/2$ supersymmetry. 
The use of the $N=2$ superconformal tensor calculus was essential in our 
construction, while we still have the vector multiplet scalars parameterizing 
a special K\"ahler manifold, and the hypermultiplet scalars parameterizing a 
quaternionic (projective) manifold or its quaternionic quotient. It is rather 
straightforward to generalize our results to the case of arbitrary 
(quaternionic) hypermultiplet couplings by using the same N=2 superconformal 
calculus and the results of ref.~\cite{calm}.  However, we cannot exclude the 
existence of a much larger class of the invariant actions with merely $N=1/2$ 
local supersymmetry, which are not derivable from the $N=2$ invariant actions 
we used. It would be interesting to find such additional invariants for a 
construction of the most general $N=1/2$ supergravity matter couplings.

We are unaware of any supergravity model to be constructed in a (curved) 
non-anticommutative superspace, so a relation of our $N=1/2$ matter-coupled 
supergravity to the non-anticommutative superspace remains unclear to us.

It may also be of interest to study the conditions of further (spontaneous) 
breaking of local $N=1$ supersymmetry by analyzing the vacuum expectation 
values of the supersymmetry transformatons of the fermionic 
fields.~\footnote{See e.g., ref.~\cite{partial}, as regards partial breaking of
 $N=2$ local supersymmetry.}

\newpage

{\large{\bf Acknowledgements}}
\vglue.2in

TH would like to thank Satoru Saito for useful discussions and encouragement.

This work is partially supported by the Japanese Society for Promotion of 
Science (JSPS) under the Grant-in-Aid programme for scientific research, and 
the bilateral German-Japanese exchange programme under the auspices of JSPS and
 DFG (Deutsche Forschungsgemeinschaft).

\vglue.3in

\appendix
\section{notation for hypermultiplets}\label{sec:not.hyper}

We follow the notation introduced in ref.~\cite{hol}. In particular,  the 
hypermultiplets \((A_i{}^\a,\z^\a)\) belong to a representation of the
 Yang-Mills group. The scalars \(A^i{}_\a\) obey a reality condition
\begin{equation}
		A^i{}_\a
		\equiv
		\Bar{A_i{}^\a}
		=
		\e^{ij}\r_{\a\b}A_j{}^\b
\end{equation}
where the matrices \(\r_{\a\b}\) are used for raising and lowering of 
greek indices. A consistency requires
\begin{equation}
		\r_{\a\b}\r^{\b\g}
		=
		-\d_\a{}^\g
	\quad
		(\r^{\a\b} \equiv \Bar{\r_{\a\b}})
\end{equation}

When writing the action, it is convenient to introduce a matrix \(d^\a{}_\b\) 
as
\begin{equation}
	d^\a{}_\b:=-\h^{[\a\g]}\r_{\g\b}
\end{equation}
where \(\h^{\a\g}\) is the real multiplication tensor in the sense 
\begin{equation}
	\h_{\a\b}=\r_{\g\a}\h^{\g\d}\r_{\d\b},
\quad
	\left( \h^{\a\b}\equiv\Bar{\h_{\a\b}} \right)
\end{equation}
The matrix \(d^\a{}_\b\) has the properties
\begin{equation}
		\Bar{d^\a{}_\b} \equiv d_\a{}^\b=d^\b{}_\a
	\quad
		{(\rm Hermitean)}~
	\qquad
		d_\a{}^\b = \r_{\g\a}\r^{\d\b}d^\g{}_\d
	\quad
		{(\rm quaternionic)}
\end{equation}

\newpage

\section{N=2 superconformal transformation laws}

In this Appendix we summarize the relevant part of the $N=2$ superconformal 
transformation laws \cite{hol} in our 2-component spinor notation, in four 
Euclidean dimensions:

(i) as regards the $N=2$ {\it Weyl\/} multiplet (physical) components,
\begin{alignat}{3}
\label{app:sc.e}
&	\d e_\m{}^a
	&=&
	-i\ve^i\s^a\Bar{\j}_{\m i}
	-i\Bar{\ve}{}_i\s^a\j_\m^i~,		\\
\label{app:sc.j^1}
&	\d \j_\m^{iA}
	&=&
	+2\bigl[
		\pa_\m\ve^{iA}
		-\frac{1}{2}\o_\m^{mn}\s_{mn}{}^A{}_B\ve^{iB}
		+\frac{1}{2}iA_\m\ve^{iA}
	\bigr]
	+\cv_\m{}^i{}_j\ve^{jA}	
\nonumber \\
& & &
	-\frac{1}{2}iT^{-ij}_{\m m}(\s^m\Bar{\ve}{}_j)^A
	+i(\s_\m\Bar{\h}{}^i)^A
	-\frac{1}{2}i\L_A\j^{iA}_\m
	+\L^i{}_j\j^{jA}_\m~,		\\
\label{app:sc.Bar{j}_2}
&	\d \Bar{\j}{}_{\m i\dot{A}}
	&=&
	+2\bigl[
		\pa_\m\Bar{\ve}{}_{i\dot{A}}
		-
		\frac{1}{2}\o_\m^{mn}\s_{mn}{}_{\dot{A}}{}^{\dot{B}}
		\Bar{\ve}{}_{i\dot{B}}
		-
		\frac{1}{2}i\Bar{A}{}_\m\Bar{\ve}{}_{i\dot{A}}
	\bigr]
	+\cv{}_\m{}_i{}^j\Bar{\ve}_{j\dot{A}}
\nonumber	\\
& & &
	+\frac{1}{2}iT^{+}_{\m mij}(\s^m\ve^j)_{\dot{A}}
	-i(\s_\m\h_i)_{\dot{A}}
	+\frac{1}{2}i\L_A\Bar{\j}{}_{\m i\dot{A}}
	+\L{}_i{}^j\Bar{\j}{}_{\m j\dot{A}}
\end{alignat}

(ii) as regards the $N=2$ {\it vector\/} multiplet components,
\begin{alignat}{3}
\label{app:sc.X}
&	\d X^I
	&=&
	-\ve^i\O^I_i
	-i\L_A X^I				\\
&	\d \Bar{X}{}^I
	&=&
	+\Bar{\ve}{}_i\Bar{\O}{}^{iI}
	+i\L_A \Bar{X}{}^I		\\
\label{app:sc.W}
&	\d W_\m^I
	&=&
	-i \Bar{\ve}{}_i \s_\m \O_j^I \ve^{ij} 
	+i \ve^i\s_\m\Bar{\O}{}^{jI}\ve_{ij}
	+2\ve^i\j_\m^j\Bar{X}^I\ve_{ij}
	-2\Bar{\ve}{}_i\Bar{\j}_{\m j}X^I\ve^{ij}~,		\\
\label{app:sc.O_2}
&	\d \O_i^{IA}
	&=&
	-2i(\s^\l\Bar{\ve}{}_i)^A
	[
		(\pa_\l+iA_\l)X^I
		-
		gf_{JK}{}^I W_\l^J X^K 
		-
		\frac{1}{2} \j^k_\l\O^I_k
	]
\nonumber 	\\
& & &
	+Y_{ij}^I\ve^{jA}
	+\s^{mnA}{}_B\cf^{-I}_{mn}\ve_{ij}\ve^{jB}
	-2gf_{JK}{}^I X^J\Bar{X}{}^K\ve_{ij}\ve^{jA}
\nonumber 	\\
& & &
	+2X^I\h_i^A
	-\frac{1}{2}i\L_A \O_i^{IA}
	+\L_i{}^j\O_j^{IA}~,			\\
&	\d \Bar{\O}{}^{iI \dot{A}}
	&=&
	-2i(\ve^i\s^\l)^{\dot{A}}
	[
		(\pa_\l-i\Bar{A}{}_\l)\Bar{X}^I
		-
		gf_{JK}{}^I W_\l^J \Bar{X}{}^K
		+
		\frac{1}{2}\Bar{\j}{}_{\l k}\Bar{\O}{}^{kI}
	]
\nonumber 	\\
& & &
	+
	Y{}^{ikI}\Bar{\ve}{}_k^{\dot{A}}
	-
	\Bar{\ve}{}_k^{\dot{B}}\s^{mn}{}_{\dot{B}}{}^{\dot{A}}\cf^{+I}_{mn}
\ve^{ik}
	-
	2gf_{JK}{}^I \Bar{X}{}^J X^K\ve^{ik}\Bar{\ve}{}_k^{\dot{A}}
\nonumber	\\
& & &
	+
	2\Bar{X}{}^I \Bar{\h}{}^{i\dot{A}}
	+
	\frac{1}{2}i\L_A \Bar{\O}{}^{iI \dot{A}}
	+
	\L^i{}_k\Bar{\O}{}^{kI \dot{A}}
\end{alignat}

(iii) and as regards {\it hyper}multiplets,
\begin{alignat}{3}
&	\d A_i{}^\a
	&=&
	-2\Bar{\z}{}^\a\Bar{\ve}{}_i
	-2\r^{\a\b}\ve_{ij}\z_\b\ve^j
	+\L_i{}^jA_j{}^\a,		\\
&	\d A^i{}_\a
	&=&
	-2\ve^i\z_\a
	-2\r_{\a\b}\ve^{ij}\Bar{\ve}{}_j\Bar{\z}{}^\b
	+\L^i{}_j A^j{}_\a,		\\
&	\d\z_{\a A}
	&=&
	+i(\Bar{\ve}{}_i\s^\m)_{A}\Bar{D}{}_\m A{}^i_\a
	+
	2gA^i{}_\b\Bar{X}{}_\a{}^\b\ve_{ij}\ve^j_A
	+
	A^i{}_\a\h_{iA}
	+
	\frac{i}{2}\L_A \z_{\a A}		\\
&	\d \Bar{\z}{}^\a_{\dot{A}}
	&=&
	+i(\s^\m\ve^i)_{\dot{A}}D_\m A_i{}^\a
	+
	2gX^\a{}_\b A_i{}^\b \ve^{ij}\Bar{\ve}{}_{j\dot{A}}
	+
	A_i{}^\a\Bar{\h}{}^i_{\dot{A}}
	-
	\frac{i}{2}\L_A \Bar{\z}{}^\a_{\dot{A}}~,
\end{alignat}
where we have used the following abbreviations:
\begin{equation}
	\cf_{\m\n}{}^I
	:=
	\hat{F}_{\m\n}{}^I-\frac{1}{4}X^I T_{\m\n ij}\ve^{ij},
\qquad
	\hat{F}_{\m\n}{}^I
	:=
	F_{\m\n}{}^I+G_{\m\n}{}^I~,	
\end{equation}
\begin{equation}
	D_\m A_i{}^\a
	:=
	\pa_\m A_i{}^\a
	+
	\frac{1}{2}\cv_{\m i}{}^j A_j{}^\a
	-
	gW_\m{}^\a{}_\b A_i{}^\b
	+
	\Bar{\z}{}^\a\Bar{\j}{}_{\m i}
	+
	\r^{\a\b}\ve_{ij}\z_\b\j_\m^j~,
\end{equation}
\begin{equation}
	\Bar{D}{}_\m A^i{}_\a
	=
	\pa_\m A^i{}_\a
	+
	\frac{1}{2}A^j{}_\a\cv_\m{}^i{}_j
	-
	gA^i{}_\b W_{\m\a}{}^\b
	+
	\j^i_\m\z_\a 
	+
	\r_{\a\b}\ve^{ij}\Bar{\j}{}_{\m j}\Bar{\z}{}^\b
\end{equation}

As for the auxiliary fields, we have e.g.,
\begin{equation}\label{app:last}
\begin{split}
	\d \cv_\m{}^i{}_j
	=
&	-3i(\ve^i \s_\m \Bar{\c}{}_j-\ve^{il}\ve_{jm}\c^m\s_\m\Bar{\ve}{}_l)
	+2\Bigl(
		\ve^i\f_{\m j}
		-\j_\m^i\h_j
		+\ve^{il}\ve_{jm}
			(
				\Bar{\ve}_l\Bar{\f}{}_\m^m
				-\Bar{\j}{}_{\m l}\Bar{\h}{}^m
			)
	\Bigr)
\\
&	-\frac{1}{2}\d^i{}_j
	\Bigl[
		-3i(\ve^k\s_\m\Bar{\c}{}_k-\c^k\s_\m\Bar{\ve}{}_k)
		+2(
			\ve^k \f_{\m k}+\Bar{\ve}{}_k\Bar{\f}{}^k_\m
			-\j_\m^k\h_k-\Bar{\j}{}_{\m k}\Bar{\h}{}^k
		)
	\Bigr]	
\end{split}
\end{equation}

\section{decomposition rules}

After the $C$-deformation (Sec.~3) and the gauge-fixing of the truly 
superconformal symmetries (Sec.~6), the $Q$-supersymmetry transformations are 
modified by the compensating $S$-supersymmetry-, chiral $U(1)$- and chiral 
$SU(2)$- transformations, whose parameters are given by 
\begin{alignat}{3}
&	\h_{i A}
	&=&
	+2g d^\a{}_\b \ve_{ij}\ve^k_A A_k{}^\g \Bar{X}{}^\b{}_\g A^j{}_\a
	+
	d^\a{}_\b \r^{\b\g}\ve_{ij} (\ve^j\s^{\r\l}){}_A
	\bigl(
		\z^\g \s_{\r\l} \z_\a
	\bigr)
\nonumber	\\
& & &
	-(\Bar{\ve}{}_j\s^\m)_A
	\Bigl[
		\frac{1}{4}N_{IJ}
		\bigl(
			\O_i^I\s_\m\Bar{\O}{}^{jJ}
			-
			\frac{1}{2}\d_i{}^j \O_k^I\s_\m\Bar{\O}{}^{kJ}
		\bigr)	
		+
		\d_i{}^j d^\a{}_\b(\z_\a \s_\m \Bar{\z}{}^\b)
	\Bigr]~,
								\\
&	\Bar{\h}{}^i_{\dot{A}}
	&=&
	+
	2g d^\a{}_\b\ve^{ij}\Bar{\ve}{}_{k{\dot{A}}} 
	A_j{}^\a X_\b{}^\g A{}^k{}_\g
	+
	d_\a{}^\b \r_{\b\g} \ve^{ij}
	(\s^{\r\l}\Bar{\ve}{}_j)_{\dot{A}}
	\bigl(
		\Bar{\z}{}^\a \s_{\r\l} \Bar{\z}{}^\g
	\bigr)
\nonumber	\\
& & &
	+
	(\s^\m \ve^j)_{\dot{A}}
	\Bigl[
		\frac{1}{4}N_{IJ}
		\bigl(
			\O_j^I\s_\m\Bar{\O}{}^{iJ}
			-
			\frac{1}{2}\d^i{}_j \O_k^I\s_\m\Bar{\O}{}^{kJ}
		\bigr)	
		+
		\d^i{}_j d_\a{}^\b(\Bar{\z}{}^\a \s_\m \z_\b)
	\Bigr],								\\
&	\L_A
	&=&
	+\frac{1}{2}iN_{IJ}
	(-\ve^i\L_i^I\Bar{X}{}^J+\Bar{\ve}{}_i\Bar{\L}{}^{iI} X^J)~, 	\\
&	\L_i{}^j
	&=&
	+2\sqrt{\frac{-2}{c}}
	\Bigl(
	d_\a{}^\b
	\bigl[
		B^j{}_\b(\Bar{\x}{}^\a\Bar{\ve}{}_i)
		-
		\frac{1}{2}\d^j{}_i 
		B^k{}_\b(\Bar{\x}{}^\a\Bar{\ve}{}_k)
	\bigr]
\nonumber	\\
& & &\qquad\qquad
	-d^\a{}_\b
	\bigl[
		B_i{}^\b(\x_\a\ve^j)
		-
		\frac{1}{2}\d_i{}^j 
		B_k{}^\b(\x_\a\ve^k)	
	\bigr]
	\Bigr)~,							
\end{alignat}
where the convenient parametrization is given by \cite{hol}
\begin{equation}
	\L_i{}^I
	:=
	\O_i^I-Z^I\O_i^0
	=
	\left(
	\begin{array}{c}
		\L_{iA}^I \\
		0 \\
	\end{array}
	\right),
\quad
	Z^I:=\fracmm{X^I}{X^0},
\quad
	\x^\a
	:=
	\z^\a-B_a{}^\a\z^a
	=
	\left(
	\begin{array}{c}
		0 \\
		\Bar{\x}{}^\a_{\dot{A}} \\
	\end{array}
	\right), \nonumber
\end{equation}
\begin{equation}
	c:=-4(A_i{}^a A^i{}_a)^{-1}=d_\a{}^\b B_a^\a B^a_\b,
\qquad
	B_a{}^\a:=A^{-1}{}^i{}_a A_i{}^\a,
\end{equation}
\begin{equation}
	a,b=1,2,
\quad
	\a,\b=3,\dots,2r.\nonumber
\end{equation}


\newpage

\begin{thebibliography}{13}
\bibitem{rev1} P. van Nieuwenhuizen, Phys. Repts. {\bf 68} (1981) 189
\bibitem{rev2}S.W. Hawking and M. Ro\v{c}ek eds., {\it Superspace and 
Supergravity}, Cambridge University Press, 1981;\\
S.J. Gates Jr., M.T. Grisaru, M. Ro\v{c}ek and W. Siegel, {\it Superspace 
or One Thousand and One Lessons in Supersymmetry}, Benjamin/Cummings Publ., 
1983;\\
A. Salam and E. Sezgin eds., {\it Supergravities in Diverse Dimensions},
Elsevier/World Scientific Publ.,  1989
\bibitem{ncgr} A.H. Chamseddine, G. Felder and J. Fr\"olich, Commun. Math. 
Phys. {\bf 155} (1993) 205 [hep-th/9209044];\\
J. Madore and J. Mourad, Int. Journ. Mod. Phys. {\bf D3} (1994) 221 
[gr-qc/9307030];\\
M. Chaichian, P.P. Kulish, K. Nishijima and A. Tureanu, Phys. Lett. {\bf B604}
(2004) 98 [hep-th/0408069];\\
P. Aschieri, C. Blohmann, M. Dimitrijevi\v{c}, F. Meyer, 
P. Schupp and J. Wess, Class. Quantum Grav. {\bf 22} (2005) [hep-th/0504183]
\bibitem{cure} H. Nishino and S. Rajpoot, Phys. Lett. {\bf B532} (2004) 334
[hep-th/0107216];\\
B.M. Zupnik, Class. and Quantum Grav. {\bf 24} (2007) 15 [hep-th/0512231]
\bibitem{cern} L. Alvarez-Gaum\'e, F. Meyer and M.A. Vazquez-Mozo,  
Nucl. Phys. {\bf B753} (2006) 92 [hep-th/0605113]
\bibitem{ov} H. Ooguri and C. Vafa, Adv. Theor. Math. Phys. {\bf 7} (2003) 
53, and {\bf 7} (2004) 405 [hep-th/0302109 and 0303063]
\bibitem{sei} N. Seiberg, JHEP {\bf 0306} (2003) 010 [hep-th/0305248]
\bibitem{our} T. Hatanaka, S.V. Ketov and S. Sasaki, Phys. Lett. {\bf B619} 
(2005) 352 [hep-th/0504191];\\
T. Hatanaka, S.V. Ketov, Y. Kobayashi and S. Sasaki, Nucl. Phys. {\bf B716}
(2005) 88, and {\bf B726} (2005) 481 [hep-th/0502026 and hep-th/0506071]
\bibitem{hol} B. de Wit, J.W. van Holten and A. van Proeyen, Nucl. Phys. 
{\bf B167} (1980) 186;\\
B. de Wit, P.G. Lauwers, R. Philippe, S.Q. Su and A. van Proeyen, Phys. Lett.
{\bf B134} (1984) 37;\\
B. de Wit and A. van Proeyen, Nucl. Phys. {\bf B245} (1984) 89;\\
B. de Wit, P.G. Lauwers and A. van Proeyen, Nucl. Phys. {\bf B255} (1985) 569
\bibitem{italy}  R. D'Auria, S. Ferrara and P. Fr\'e, Nucl. Phys. {\bf B359} 
(1991) 705; \\
L. Andrianopoli, M. Bertolini, A. Ceresole, R. D'Auria,
S. Ferrara and P. Fr\'e, Nucl. Phys. {\bf B476} (1996) 397 [hep-th/9603004]
\bibitem{hat}  T. Hatanaka and S.V. Ketov, Class. and Quantum Grav.
{\bf 23} (2006) L45 [hep-th/0602115]
\bibitem{fn} S. Ferrara and  P. van Nieuwenhuizen, Phys. Rev. Lett. {\bf 37}
(1976) 1669
\bibitem{2to1} L. Andrianopoli, R. D'Auria and S. Ferrara, Nucl. Phys. 
{\bf B628} (2002) 387 [hep-th/0112192]
\bibitem{apro} A. van Proeyen, {\it Superconformal Tensor Calculus In N=1 And 
N=2 Supergravity}, in `Supersymmetry and Supergravity 1983', World Sci., 1983, 
p.~578  
\bibitem{tnw} P. van Nieuwenhuizen and A. Waldron, Phys. Lett. {\bf B389} 
(1996) 29 [hep-th/9608174], and {\it A Continuous Wick rotation for spinor
 fields and supersymmetry in Euclidean space}, in `Gauge theories, Applied 
Supersymmetry and Quantum Gravity', London, 1996, p.~394 [hep-th/9611043];\\
U. Theis and P. van Nieuwenhuizen, Class. Quantum Grav. {\bf 18} (2001) 5469 
[hep-th/0108204]
\bibitem{calm} B. de Wit, B. Kleijn and S. Vandoren, Nucl. Phys. {\bf B568}
(2000) 475 [hep-th/9909228]
\bibitem{partial} S. Cecotti, L. Girardello and M. Porrati, Phys. Lett. 
{\bf B168} (1986) 83;\\
S. Ferrara, L. Girardello and M. Porrati, Phys. Lett. {\bf B366} (1996) 155 
[hep-th/9510074], and {\bf B376} (1996) 275 [hep-th/9512180];\\
L. Andrianopoli, R. D'Auria, S. Ferrara and M.A. Lledo, JHEP {\bf 0301} (2003) 
045 [hep-th/0212236];\\
H. Itoyama and K. Maruyoshi, Int. J. Mod. Phys. {\bf A21} (2006) 6191 
[hep-th/0603180]. 

\end{thebibliography}
\end{document}